\documentclass[aps,preprint,amsmath,amssymb,superscriptaddress,nofootinbib]{revtex4}

\usepackage{graphicx}
\begin{document}

\title{A hypothesis on neutrino helicity}

\author{\.{I}. \c{S}ahin}
\email[]{inancsahin@ankara.edu.tr}
 \affiliation{Department of
Physics, Faculty of Sciences, Ankara University, 06100 Tandogan,
Ankara, Turkey}

\begin{abstract}
It is firmly established by experimental results that neutrinos are
almost $100\%$ longitudinally polarized and left-handed. It is also
confirmed by neutrino oscillation experiments that neutrinos have
tiny but non-zero masses. Since their masses are non-zero the
neutrinos cannot be strictly described by pure helicity states which
coincide with the chirality eigenstates. On the other hand, it is
generally assumed that ultrarelativistic massive fermions can be
described well enough by the Weyl equations. This assumption
obviously explains why the neutrinos are almost 100\% longitudinally
polarized.  We discuss the validity of this assumption and show that
the assumption is fallacious for a fermion with a general spin
orientation. For instance, a fermion with a transverse polarization
(relative to its momentum) cannot be described by one of the Weyl
equations even in the ultrarelativistic limit. Hence, the fact that
neutrinos are almost completely longitudinally polarized cannot be
explained in the basis of relativistic quantum mechanics or quantum
field theory. As a solution to this problem, we propose a new
hypothesis according to which neutrinos are strictly described by
pure helicity states although they are not massless.

\end{abstract}


\maketitle

\section{Introduction}
In the original version of the Standard Model of particle physics
neutrinos are accepted to be massless. Hence, they are described by
pure helicity states which is consistent with the results obtained
from experiments
\cite{Goldhaber,Roesch:1982ry,Fetscher:1984da,Heister:2001me,Abe:1997dy}.
On the other hand, neutrino oscillation experiments point out an
extension of the Standard Model. In the minimal extension of the
Standard Model with massive neutrinos, flavor and mass eigenstates
do not coincide. Flavor eigenstates can be written as a
superposition of mass eigenstates through the mixing equation
$\nu_{\ell L}=\sum_{i=1}^3U_{\ell i}\;\nu_{i L}$  where  $U_{\ell
i}$ is the Pontecorvo-Maki-Nakagawa-Sakata (PMNS) matrix element
\cite{Bilenky}. Since a flavor eigenstate is a mixture of different
mass eigenstates each has a different rest frame, the rest frame of
a flavor neutrino is somewhat uncertain.\footnote{In this paper only
Dirac neutrinos have been considered. According to Dirac equation
the states of definite momentum are not eigenstates of velocity.
Hence, one may argue that the rest frame for a massive fermion is
always uncertain. On the other hand, if we perform a time average
over the period of Zitterbewegung (which is an extremely small time
period) we obtain the classical velocity $\frac{c^2\vec p}{E}$. In
the case of flavor neutrinos we have an additional uncertainty in
the rest frame due to neutrino mixing and this uncertainty cannot be
removed even though an average over the period of Zitterbewegung is
performed.} However, the notion of spin is closely related to the
rest frame of the particle. The spin four-vector for a fermion
\begin{eqnarray}
\label{fourspin}
s^\mu=\left(\frac{\vec{p}\cdot\vec{s'}}{m},\vec{s'}+\frac{\vec{p}\cdot\vec{s'}}{m(E+m)}\vec{p}\right)
\end{eqnarray}
is obtained by a Lorentz boost of $(s^\mu)_{RF}=(0,\vec{s'})$ from
the rest frame of the particle \cite{GreinerQED}. Therefore it is
reasonable to accept that the spin state for a flavor neutrino is
not well-defined. As far as we know, this problem has been skipped
in the literature possibly because neutrino masses are extremely
small ($m_{1,2,3}\lesssim 1\; eV$), and hence it is a very good
approximation to use expressions obtained in the limit $m_{\nu_i}\to
0$. It is generally believed that free solutions of the Dirac
equation coincide with pure helicity states (which correspond to
chirality eigenstates) in the zero-mass limit.\footnote{ We should
clarify the terminology used in this paper. Sometimes, when the spin
three-vector of a fermion with non-zero mass is oriented parallel
(anti-parallel) to the direction of momentum, the corresponding
spinor is called right-handed (left-handed). But since you can
convert a right-handed fermion into a left-handed one simply by
changing your frame of reference, the helicity states defined in
this way are not intrinsic to the particle. When we use the term
left(right)-handed helicity for a fermion with non-zero mass we
imply the above meaning of the helicity. On the other hand, when we
use the term {\it pure helicity states} we imply Lorentz invariant
helicity for a massless fermion which correspond to chirality
eigenstates. Of course, two different meaning of the helicity
coincide in the $m\to0$ limit.} Therefore one can assume that each
of the constituent mass eigenstates for a flavor neutrino is
described well enough by a pure helicity state. It is natural to
define the spin of a flavor neutrino in terms of the spin of its
constituent mass eigenstates. Hence, one may conceive the spin state
for a flavor neutrino as a superposition of the spin states of
different mass eigenstates. Since the spin of all mass eigenstates
are equal to each other with a high accuracy (they are all
approximately left-handed) we can uniquely define the spin of the
flavor eigenstate. Here, we should be aware of the approximation
that is used in the definition of the spin for a flavor eigenstate.
Strictly speaking, the notion of spin and its special orientation
helicity (see footnote 2) is not uniquely defined for a flavor
neutrino. The assumption that a flavor neutrino has a uniquely
defined spin state that coincides with left-handed helicity is an
approximation which is valid with some degree of accuracy. This
approximation is based on the assumption that ultrarelativistic
massive fermions can be described well enough by the Weyl equations.
According to this assumption, the free solutions of the Dirac
equation containing a very small mass term (compared to the energy
scale that we consider) can be described by pure helicity states
with a high degree of accuracy. On the contrary, in the next section
we will discuss some counter evidences obtained from spin dependent
neutrino cross section and relativistic quantum mechanics that show
this assumption is not valid in general. In the absence of this
assumption, we encounter some serious problems. We cannot define the
spin state of a flavor neutrino uniquely. We can only conceive the
flavor spin as a superposition of the spin states of its constituent
mass eigenstates where each spin state may have an arbitrary spin
orientation. This result obviously contradicts with firmly
established experimental results that show flavor neutrinos are
almost $100\%$ longitudinally polarized and left-handed. We call
this problem the neutrino helicity problem. Our new hypothesis is an
attempt which is proposed as a solution to the neutrino helicity
problem.

The organization of the paper is as follows. In the next section
first we will introduce our new hypothesis and then discuss some
evidences obtained from spin dependent cross section and
relativistic quantum mechanics that show generally believed
assumption discussed in the previous paragraph is not valid in
general. Absence of this assumption leads to the neutrino helicity
problem and forces us to assert a new hypothesis. In the conclusion
section we will discuss the implications of the new hypothesis on
the foundations of physics.

\section{A hypothesis on neutrino helicity and some evidences}

In his famous $1939$ paper Wigner investigated unitary
representations of the Poincar\'{e} group and classified particles
according to their internal space-time symmetries \cite{Wigner}. One
of the important criteria used in Wigner's classification is the
existence of the rest frame of a particle. There is no frame of
reference in which a massless particle such as a photon is at rest.
Hence the little group for a massless particle is $E(2)$-like. On
the other hand, a massive particle has a rest frame and in this
frame we can rotate its spin three-vector without changing the
momentum. Its little group is then $O(3)$-like.  We assert the
following hypothesis which makes a flavor neutrino $100\%$
longitudinally polarized: The neutrino flavor eigenstate is a
mixture of different mass eigenstates each has a different rest
frame. Hence, the rest frame of a flavor neutrino is uncertain and
this uncertainty makes its rest frame undefined. Since there is no
frame of reference in which a flavor neutrino is at rest, its little
group is no more $O(3)$-like. It should be classified together with
massless particles described by $E(2)$-like little group, although
it has a non-zero mass and does not propagate at the speed of light.
As we have discussed in the introduction, one may interpret the spin
state for a flavor neutrino as a superposition of the spin states of
different mass eigenstates. We think that this interpretation is not
correct. Our interpretation is the following: since the definition
of spin requires the existence of the rest frame and the rest frame
of a flavor neutrino is uncertain, we do not have a well-defined
spin state for a flavor neutrino. On the other hand, the definition
of helicity (we mean a Lorentz invariant helicity which coincides
with chirality. Please see footnote 2.) does not require the
existence of the rest frame. Consequently, the helicity of the
flavor neutrino is a well-defined quantity.

\subsection{Evidences From Spin Dependent Cross
section}

It is generally assumed that ultrarelativistic massive fermions can
be described well enough by the Weyl equations. Hence neutrinos are
accepted to be completely longitudinally polarized and pure helicity
states for the neutrino fields are used in the cross section
calculations. Let us perform cross section calculations for
neutrinos with a general spin orientation and probe the validity of
this assumption. We will consider a simple particular process,
namely polarized neutrino production via electron capture, where
much of the computation can be done easily. This process can be
written at the quark level as $e^- u \to \nu_i d$, where $\nu_i$
represents a neutrino in the mass eigenstate. The process $e^- u \to
\nu_i d$ is described by a t-channel $W$ exchange diagram. Spin
dependent amplitude for the process is given by\footnote{We assume
that the $W$ propagator can be approximated as
$\frac{(g_{\mu\nu}-q_\mu
q_\nu/m_w^2)}{q^2-m_w^2}\approx-\frac{g_{\mu\nu}}{m_w^2}$.}
\begin{eqnarray}
\label{amplitude1}
M=\frac{G_F}{\sqrt{2}}U_{ei}U_{ud}\left[\bar{u}(p_{\nu_i},s'_{\nu_i})\hat{\Sigma}(s_{\nu_i})\gamma^\mu(1-\gamma_5)u(p_e,s'_e)\right]
\left[\bar{u}(p_d,s'_d)\gamma_\mu(1-\gamma_5)u(p_u,s'_u)\right]
\end{eqnarray}
where $G_F$ is the Fermi constant, $U_{ei}$ is the PMNS matrix
element, $U_{ud}$ is the Cabibbo-Kobayashi-Maskawa (CKM) matrix
element and $\hat{\Sigma}(s_{\nu_i})\equiv\frac{1}{2}(1+\gamma_5
\gamma_\mu s_{\nu_i}^\mu)$ is the covariant spin projection operator
for the neutrino. In the rest frame of the neutrino the spin
four-vector is $(s_{\nu_i}^\mu)_{RF}=(0,\vec{s}_{\nu_i})$. In an
arbitrary reference frame the spin four-vector can be obtained by a
Lorentz boost from the rest frame. In a reference frame where the
neutrino has a momentum $\vec p$ and energy $E$ its spin four-vector
can be defined by Eq.(\ref{fourspin}) with $m=m_{\nu_i}$ and
$\vec{s'}=\vec{s}_{\nu_i}$. When we square the amplitude and sum
over fermion spins the projection $\hat{\Sigma}(s_{\nu_i})
u(p_{\nu_i},s'_{\nu_i})=\delta_{{s'_{\nu_i}},{s_{\nu_i}}}
u(p_{\nu_i},s_{\nu_i})$ ensures that the sum over $s'_{\nu_i}$
yields just one term with $s'_{\nu_i}=s_{\nu_i}$ \cite{GreinerQED}.
The spin-summed squared amplitude is calculated to be
\begin{eqnarray}
\label{amplitudesq1} \sum_{s'_{\nu_i},s'_e,s'_d,s'_u}|M|^2=&&64
G_F^2 |U_{ei}|^2 |U_{ud}|^2 \left[(p_e\cdot p_u)(p_{\nu_i}\cdot
p_d)-m_{\nu_i}(p_e\cdot p_u)(s_{\nu_i}\cdot p_d)\right].
\end{eqnarray}
It describes polarized neutrinos with spin four-vector $s_{\nu_i}$
but unpolarized electrons, u and d quarks. At first glance, it seems
as if spin dependent term in Eq.(\ref{amplitudesq1}) vanishes in the
$m_{\nu_i}\to 0$ limit. But spin four-vector contains terms
inversely proportional to $m_{\nu_i}$. Therefore first we should
perform Lorentz scalar products and then examine its zero-mass
limit. After Lorentz scalar products are performed, the spin-summed
squared amplitude can be written as
\begin{eqnarray}
\label{amplitudesq2} \sum_{s'_{\nu_i},s'_e,s'_d,s'_u}|M|^2=&&64
G_F^2 |U_{ei}|^2 |U_{ud}|^2 (E_eE_u-\vec p_e\cdot \vec p_u)
(E_{\nu_i}E_d-E_d (\vec s_{\nu_i}\cdot \vec p_{\nu_i})\nonumber \\
&&+\vec p_{\nu_i}\cdot \vec p_d\left(\frac{\vec s_{\nu_i}\cdot \vec
p_{\nu_i}}{E_{\nu_i}+m_{\nu_i}}-1\right)+m_{\nu_i}\vec
s_{\nu_i}\cdot \vec p_d).
\end{eqnarray}
We observe from Eq.(\ref{amplitudesq2}) that the term $\vec
s_{\nu_i}\cdot \vec p_{\nu_i}$ does not completely vanish in the
$m_{\nu_i}\to 0$ limit. Therefore it doesn't matter how small it is,
if the neutrino has a nonzero mass then the cross section depends on
its spin orientation. In the center-of-momentum frame zero-mass
limit of the squared amplitude becomes
\begin{eqnarray}
\label{limit1} \lim_{m_{\nu_i}\to
0}\sum_{s'_{\nu_i},s'_e,s'_d,s'_u}|M|^2=&&64 G_F^2 |U_{ei}|^2
|U_{ud}|^2 (E_eE_u+|\vec p_e||\vec p_u|)\nonumber
\\&&\times(E_d+|\vec p_d|)(E_{\nu_i}-\vec s_{\nu_i}\cdot \vec
p_{\nu_i}).
\end{eqnarray}
We see from the above expression that the squared amplitude and
hence the cross section takes its largest value when the neutrino is
left-handed ($\vec s_{\nu_i}=-\frac{\vec p_{\nu_i}}{|\vec
p_{\nu_i}|}$) and zero when the neutrino is right-handed ($\vec
s_{\nu_i}=+\frac{\vec p_{\nu_i}}{|\vec p_{\nu_i}|}$). If the spin
three-vector is perpendicular to the direction of neutrino momentum
then the cross section is half of the cross section for left-handed
neutrino. In general, if the spin three-vector makes an angle $\phi$
with respect to direction of neutrino momentum then we deduce that
the zero-mass limits of the cross sections in the center-of-momentum
frame for spin up and spin down polarizations are given by
\begin{eqnarray}
\label{crosssection1}
\sigma^{(\uparrow)}(s_{\nu_i})=\sin^2\left(\frac{\phi}{2}\right)\sigma^{(L)}\\
\label{crosssection2}
\sigma^{(\downarrow)}(s_{\nu_i})=\cos^2\left(\frac{\phi}{2}\right)\sigma^{(L)}
\end{eqnarray}
where $\sigma^{(L)}$ represents cross section for left-handed
neutrino. (Of course, unpolarized total cross section remains
unchanged, i.e.,
$\sigma_{unpol}=\sigma^{(\uparrow)}+\sigma^{(\downarrow)}=\left[\sin^2\left(\frac{\phi}{2}\right)+\cos^2\left(\frac{\phi}{2}\right)\right]\sigma^{(L)}=\sigma^{(L)}$)
In the above equations the limit $m_{\nu_i}\to 0$ is implemented but
not shown. We should note that spin down polarization corresponds to
a spin three-vector which is oriented opposite to the direction of
spin three-vector for spin up, i.e., spin three-vector for spin down
polarization makes an angle $\phi+\pi$ with respect to direction of
momentum. In opposition to expectations these results indicate that
the transverse component of the spin three-vector of a fermion does
not vanish in the zero-mass limit.\footnote {When we use the
notation $m\to 0$ or the phrase "zero-mass limit" we mean an
infinitesimal mass but not equal to zero.} This result contradicts
with the generally accepted assumption that when the speed of a
particle approaches the speed of light its spin vector lays down on
the momentum direction. This assumption is obviously related to the
assumption mentioned in the introduction section which says free
solutions of the Dirac equation coincide with pure helicity states
in the zero-mass limit. Is it indeed possible that this generally
accepted assumption is fallacious? We should examine what the
relativistic quantum mechanics says about the zero-mass limit of a
spinor describing a general spin orientation.

\subsection{Evidences From Relativistic Quantum Mechanics}

Let us construct the spinors describing a general spin orientation
and examine their behavior in the zero-mass limit. Assume that $S$
defined by spatial axes $x$-$y$-$z$ and time $t$ is the rest frame
of a spin-$1/2$ particle with mass $m$. In the rest frame of the
particle we can safely use Pauli spinors to define its spin. Let
$q$-axis be the spin quantization axis in the $z$-$x$ plane which
makes an angle $\phi$ with respect to $z$-axis. Then the
non-relativistic $2\times2$ spin matrix is
$\hat{S}=\frac{1}{2}(\sin\phi\; \hat\sigma_x+\cos\phi\;
\hat\sigma_z)$ where $\hat\sigma_x,\hat\sigma_y$ and $\hat\sigma_z$
are Pauli spin matrices. The eigenvectors of the spin matrix are
\begin{eqnarray}
\chi_+=\left(
           \begin{array}{c}
              \cos\phi/2\\
              \sin\phi/2\\
           \end{array}
         \right) \;\;\;\;\;\;\;\;
         \chi_-=\left(
           \begin{array}{c}
              -\sin\phi/2\\
              \cos\phi/2\\
           \end{array}
         \right).
\end{eqnarray}
Here $\chi_+$ and $\chi_-$ are the eigenvectors which correspond to
eigenvalues $\lambda=+1$ and $\lambda=-1$ respectively, i.e.,
$\hat{S}\chi_+=\chi_+$ and $\hat{S}\chi_-=-\chi_-$. We have so far
considered only $2\times1$ matrix representations. In the Weyl
representation, $4\times1$ Dirac spinors for a spin-$1/2$ particle
can be written as \cite{Ryder}
\begin{eqnarray}
u(p)=\left(\begin{array}{c}
\phi_R(p) \\
\phi_L(p) \\
\end{array}\right)
\end{eqnarray}
where $\phi_R(p)$ and $\phi_L(p)$ are $2\times1$ spinors and
subscripts "R" and "L" represents chirality. In the rest frame of
the particle the equality $\phi_R(0)=\phi_L(0)$ holds. $\phi_R(0)$
and $\phi_L(0)$ can be considered as the eigenvectors of the
$2\times2$ spin matrix. The spinor $u(p)$ for a particle with
four-momentum $p^\mu=(E,\vec{p})$ can be obtained via Lorentz boost
from rest spinor $u(0)$. Suppose that $S'$ frame is moving along the
negative z-axis with relative speed $v$ with respect to $S$. If we
choose $\phi_R(0)=\phi_L(0)=\chi_+$ and perform a Lorentz boost into
$S'$ frame then we obtain a Dirac spinor which describes a particle
of four-momentum $p^\mu=(E,0,0,p_z)$ and four-spin given in
Eq.(\ref{fourspin}) where $\vec{s'}=\sin\phi\; \hat x+ \cos\phi\;
\hat z$ is the unit vector on the spin quantization axis $q$ in the
rest frame of the particle. The spinor obtained in this way
describes a "spin up" state. Similarly if we choose
$\phi_R(0)=\phi_L(0)=\chi_-$ and perform a Lorentz boost, then we
obtain the "spin down" spinor.\footnote{Hereafter we will use the
notation $u(p,s)$ instead of $u(p)$ and sometimes use the
superscripts $\uparrow$ and $\downarrow$ for spin up and spin down
and the superscripts R and L for right-handed and left-handed.}
After some straightforward calculations, these "spin up"
$(\uparrow)$ and "spin down" $(\downarrow)$ spinors are found to be
\begin{eqnarray}
\label{spinup}
u^{(\uparrow)}(p,s)=\cos{\left(\frac{\phi}{2}\right)}\;u^{(R)}(p,s_*)+\sin{\left(\frac{\phi}{2}\right)}\;u^{(L)}(p,s_*)\\
\label{spindown}
u^{(\downarrow)}(p,s)=\cos{\left(\frac{\phi}{2}\right)}\;u^{(L)}(p,s_*)-\sin{\left(\frac{\phi}{2}\right)}\;u^{(R)}(p,s_*)
\end{eqnarray}
where $u^{(R)}(p,s_*)$ and $u^{(L)}(p,s_*)$ represent right-handed
and left-handed spinors, i.e., $\vec{s_*'}=\lambda\frac{\vec
p}{|\vec p|}$, $\lambda=+1(-1)$ for right-handed (left-handed). We
observe from Eqs. (\ref{spinup}) and (\ref{spindown}) that "spin up"
and "spin down" spinors can be written as a superposition of right-
and left-handed spinors. We now come to a controversial point. One
may assume that the angle $\phi$ converges to zero due to
relativistic aberration when the relative speed approaches the speed
of light. We claim that this assumption is fallacious because $\phi$
is the angle measured in the frame in which the particle is at rest.
In analogy with the term proper time we can call it "proper angle".
Then, coefficients $\cos{\left(\frac{\phi}{2}\right)}$ and
$\sin{\left(\frac{\phi}{2}\right)}$ do not depend on the energy or
the mass of the particle. Hence, these "spin up" and "spin down"
spinors do not approach one type of helicity state (R or L) when the
mass approaches zero. They are always given by the same
superposition of right- and left-handed spinors. For the special
case $\phi=\pi/2$, spin three-vector $\vec{s'}$ becomes
perpendicular to the direction of momentum. In this case,
$u^{(\uparrow)}(p,s)$ can be considered as a mixed state composed of
helicity states where each helicity state has equal probability.
(The similar thing is also true for $u^{(\downarrow)}(p,s)$.) This
mixed state remains intact for every value of the mass greater than,
but not equal to zero, i.e., $m>0$. Since the helicity and chirality
states coincide in the ultrarelativistic limit we can deduce that
the spinors $u^{(\uparrow)}(p,s)$ and $u^{(\downarrow)}(p,s)$ do not
converge to a pure helicity or a chirality state.

We conclude from the above discussion that although the helicity
states converge to the chirality eigenstates  when $m\rightarrow 0$,
the free solutions of Dirac equation describing a general spin
orientation {\it do not continuously} converge to chirality
eigenstates. Here the phrase "{\it do not continuously}" is used to
indicate the discontinuity at $m=0$. We have shown that when the
mass parameter is in the open interval $(0,\infty)$  general spin
states and chirality eigenstates are disjointed. But in the point
$m=0$ the solution describing a general spin orientation jumps to a
pure helicity state. This behavior seems contradictory to the fact
that the $E(2)$-like little group is the Lorentz-boosted $O(3)$-like
($SU(2)$-like) little group for massive particles in the zero-mass
limit \cite{Kim1,Kim2,Kim3}. However, the group contraction is
established by relating the generators of these groups and hence it
points out a local isomorphism between $SU(2)$ and $E(2)$ in the
limit $m\to 0$. A general spin orientation can be obtained by a {\it
finite} rotation from the direction in which the Lorentz boost is
performed. Therefore it is controversial to conclude from a local
isomorphism that the transverse component of the spin three-vector
vanishes in the zero-mass limit. One may observe from
Eq.(\ref{fourspin}) that when the speed of a particle approaches the
speed of light, spatial component of its spin four-vector lays down
on the momentum direction. Such a relativistic aberration is indeed
true but the spatial component of the spin four-vector does not
represent the spin orientation of the moving fermion. The true spin
three-vector should be the one which is attached on the moving frame
and coincides with the spin quantization axis. Therefore it should
be: $\vec s-\vec
vt=\vec{s'}-\frac{\vec{p}\cdot\vec{s'}}{E(E+m)}\vec{p}$. It
corresponds to an imaginary ruler on the moving frame that coincides
with the spin quantization axis. It is obvious that a ruler which is
oriented transverse to the direction of Lorentz boost remains
unchanged after the Lorentz transformation.

Cross section calculations can also be done directly by inserting
the corresponding explicit expressions for Dirac matrices and
spinors. But of course in this case, it is not necessary to use spin
projection operator $\hat{\Sigma}(s_{\nu_i})$. It can be omitted
from the amplitude. The standard model neutrinos couple minimally to
other standard model particles only through $V-A$ type vertex and
hence the interaction project out the "left" chiral component of the
field and, in the $m\to 0$ limit, the "right" component decouples
completely. But we see from Eqs. (\ref{spinup}) and (\ref{spindown})
that the spinors describing a general spin orientation (specifically
transverse polarization) cannot be separated into left and right
chirality eigenstates even in the $m\to 0$ limit. Hence the left
chirality projection operator $\hat L=\frac{1}{2}(1-\gamma_5)$
annihilates only right-handed constituent of the spinor whereas its
left-handed constituent remains intact. To be precise, if we apply
left chirality projection operator to "spin-up" and "spin-down"
spinors in Eqs. (\ref{spinup}) and (\ref{spindown}) we obtain
\begin{eqnarray}
\label{chiralityprojection} \hat
Lu^{(\uparrow)}=\sin{\left(\frac{\phi}{2}\right)}\;u^{(L)},\;\;\;\;
\hat Lu^{(\downarrow)}=\cos{\left(\frac{\phi}{2}\right)}\;u^{(L)}
\end{eqnarray}
in the $m\to 0$ limit. For $\phi=\frac {\pi}{2}$ (transverse
polarization) these give $\hat Lu^{(\uparrow)}=\hat
Lu^{(\downarrow)}=\frac{1}{\sqrt 2} u^{(L)}$. It is easy to verify
cross section formulas given in Eqs. (\ref{crosssection1}) and
(\ref{crosssection2}) using identities in
Eq.(\ref{chiralityprojection}). Therefore, two different calculation
techniques -the first technique is based on the use of covariant
spin projection operator $\hat{\Sigma}(s_{\nu_i})$ and the second
one is the direct calculation using explicit expressions for Dirac
matrices and spinors- give exactly the same results for the process
$e^- u \to \nu_i d$ with a general neutrino spin, i.e., for an
arbitrary $\phi$ value. Consequently, relativistic quantum mechanics
verifies the results obtained from spin dependent cross section
calculations. Moreover, with the help of Dirac spinors describing a
general spin orientation we deduce that the behavior of the spin
dependent cross section in the zero-mass limit is not peculiar to
the particular process $e^- u \to \nu_i d$ instead, all standard
model processes where neutrinos take part in the initial or final
states exhibit such a behavior. This is evident from the identities
given in Eq.(\ref{chiralityprojection}) and the $V-A$ coupling of
the standard model neutrinos.

\subsection{Discussion}

Both spin dependent cross section calculations and the results
obtained from relativistic quantum mechanics provide convincing
evidences against the generally accepted assumption that
ultrarelativistic massive fermions can be described well enough by
the Weyl equations. We think that the evidences are convincing
enough to reject the assumption. We have deduced that the spinors
describing a general spin orientation (specifically transverse
polarization) cannot be separated into left and right chirality
eigenstates even in the zero-mass limit. Hence, a fermion with a
general spin orientation cannot be described by one of the Weyl
equations even though it possesses an extremely small mass.
Consequently, the polarized cross section for producing a neutrino
mass eigenstate with a general spin orientation is not small.
Specifically, if the neutrino is right-handed the polarized cross
section is almost zero (it is strictly zero for $m_i\to0$) but if
the neutrino is transversely polarized (relative to its momentum)
the polarized cross section is almost half of the cross section for
left-handed neutrino (it is strictly half of the left-handed cross
section for $m_i\to0$). Therefore, the probability of producing a
neutrino mass eigenstate with a spin orientation different from
left-handed polarization is not small. Then we can conceive the
flavor eigenstate as a superposition of the mass eigenstates where
each mass eigenstate may have an arbitrary spin orientation which
can be very different from left-handed polarization. On the
contrary, experimental results confirm the fact that flavor
neutrinos are almost $100\%$ longitudinally polarized and
left-handed. Therefore, we cannot explain these experimental results
based solely on relativistic quantum mechanics or quantum field
theory. This lack of explanation is the neutrino helicity problem
that we have mentioned in the introduction section. The generally
accepted assumption that we have falsified provided a fake solution
to this problem. Hence, a new assumption is necessary to explain
experimental results. We think that the simplest solution to the
neutrino helicity problem is the new hypothesis that we have
proposed. Here, we should draw attention to the following point. Our
hypothesis and also the generally accepted assumption that we have
falsified, are not about the left-handedness of neutrinos. Our
hypothesis is an attempt to explain why the flavor neutrinos are
completely longitudinally polarized ,i.e., transverse polarization
vanishes for flavor neutrinos. The fact that all neutrinos are
left-handed but all anti-neutrinos are right-handed is out of the
scope of the hypothesis.

\section{Conclusions}
Massive particles which do not have a rest frame were not considered
in the Wigner's work \cite{Wigner}. On the other hand, quantum
mechanics makes such a peculiar case possible. Based on the
discussion about neutrino spin we conclude that a modification or an
extension of the Wigner's work is necessary. This modification is
probably related to a more deeper problem, which is the unification
of quantum mechanics with special relativity. It is generally
believed that the unification of quantum mechanics with special
relativity has been completed. Their offspring is the quantum field
theory. Nevertheless, a new hypothesis and its evidences discussed
in this paper raise some doubt on the completeness of this
unification.

\end{document}